\documentclass[]{elsarticle}

\usepackage[utf8]{inputenc}       
\usepackage[T1]{fontenc}          

\usepackage{booktabs}

\usepackage{siunitx}
\newcommand\unit[3][]{\SI[#1]{#2}{#3}}

\usepackage{color}
\usepackage{amsmath}
\usepackage{amssymb}
\usepackage{mathrsfs}             

\usepackage{lineno}

\usepackage{tikz}
\usetikzlibrary{plotmarks,patterns,calc,decorations.markings,arrows,
                decorations.pathmorphing,intersections}

\definecolor{lightblue}{rgb}{0.247059,0.407843,1}
\definecolor{darkred}{rgb}{0.5,0,0}

\usepackage[format=plain,margin=0.8cm,font=small,labelfont=bf]{caption}
\newcommand{\widecapt}{\captionsetup{margin=0.2cm}}

\newlength{\smfigwidth}
\newlength{\figwidth}
\newlength{\captwidth}
\usepackage[pdflatex]{trimclip}

\def\TickLSize{2}
\def\AxisLSize{2.2}

\title{An algorithm for the reconstruction of neutrino-induced showers
       in the ANTARES neutrino telescope}
\author[Colmar]{A.~Albert}
\author[UPC]{M.~Andr\'e}
\author[Genova]{M.~Anghinolfi}
\author[Erlangen]{G.~Anton}
\author[UPV]{M.~Ardid}
\author[CPPM]{J.-J.~Aubert}
\author[APC]{T.~Avgitas}
\author[APC]{B.~Baret}
\author[IFIC]{J.~Barrios-Mart\'{\i}}
\author[LAM]{S.~Basa}
% ...

\address[Colmar]{\scriptsize{GRPHE - Universit\'e de Haute Alsace - Institut universitaire de technologie de Colmar, 34 rue du Grillenbreit BP 50568 - 68008 Colmar, France}}
\address[UPC]{\scriptsize{Technical University of Catalonia, Laboratory of Applied Bioacoustics, Rambla Exposici\'o, 08800 Vilanova i la Geltr\'u, Barcelona, Spain}}
\address[Genova]{\scriptsize{INFN - Sezione di Genova, Via Dodecaneso 33, 16146 Genova, Italy}}
\address[Erlangen]{\scriptsize{Friedrich-Alexander-Universit\"at Erlangen-N\"urnberg, Erlangen Centre for Astroparticle Physics, Erwin-Rommel-Str. 1, 91058 Erlangen, Germany}}
\address[UPV]{\scriptsize{Institut d'Investigaci\'o per a la Gesti\'o Integrada de les Zones Costaneres (IGIC) - Universitat Polit\`ecnica de Val\`encia. C/  Paranimf 1, 46730 Gandia, Spain}}
\address[CPPM]{\scriptsize{Aix Marseille Univ, CNRS/IN2P3, CPPM, Marseille, France}}
\address[APC]{\scriptsize{APC, Univ Paris Diderot, CNRS/IN2P3, CEA/Irfu, Obs de Paris, Sorbonne Paris Cit\'e, France}}
\address[IFIC]{\scriptsize{IFIC - Instituto de F\'isica Corpuscular (CSIC - Universitat de Val\`encia) c/ Catedr\'atico Jos\'e Beltr\'an, 2 E-46980 Paterna, Valencia, Spain}}
\address[LAM]{\scriptsize{LAM - Laboratoire d'Astrophysique de Marseille, P\^ole de l'\'Etoile Site de Ch\^ateau-Gombert, rue Fr\'ed\'eric Joliot-Curie 38,  13388 Marseille Cedex 13, France}}
% ...
\address[Genova-UNI]{\scriptsize{Dipartimento di Fisica dell'Universit\`a, Via Dodecaneso 33, 16146 Genova, Italy}}
% ...

\makeatletter
\usepackage[pdftitle={\@title}, pdfauthor={Tino Michael}, bookmarks=true,
            pdfborder={0 0 0}]{hyperref}
\usepackage[all]{hypcap}          

\usepackage{scrextend}

\journal{ApJ}

\begin{document}


\begin{frontmatter}


\begin{abstract}
    Muons created by $\nu_\mu$ charged current (CC) interactions in the water surrounding
    the ANTARES neutrino telescope have been almost exclusively used so far in searches
    for cosmic neutrino sources. Due to their long range, highly energetic muons inducing
    Cherenkov radiation in the water are reconstructed with dedicated algorithms that
    allow the determination of the parent neutrino direction with a median angular
    resolution of about \unit{0.4}{\degree} for an $E^{-2}$ neutrino spectrum.
    In this paper, an algorithm optimised for accurate reconstruction of energy and
    direction of shower events in the ANTARES detector is presented.
    Hadronic showers of electrically charged particles are produced by the disintegration
    of the nucleus both in CC and neutral current (NC) interactions of neutrinos in water.
    In addition, electromagnetic showers result from the CC interactions of electron
    neutrinos while the decay of a tau lepton produced in $\nu_\tau$ CC interactions will
    in most cases lead to either a hadronic or an electromagnetic shower.
    A shower can be approximated as a point source of photons. With the presented method,
    the shower position is reconstructed with a precision of about \unit{1}{\metre}; the
    neutrino direction is reconstructed with a median angular resolution between
    \unit{2}{\degree} and \unit{3}{\degree} in the energy range of \SIrange{1}{1000}{TeV}.
    In this energy interval, the uncertainty on the reconstructed neutrino energy is about
    \SIrange{5}{10}{\%}. The increase in the detector sensitivity due to the use of
    additional information from shower events in the searches for a cosmic neutrino flux
    is also presented.
\end{abstract}

\end{frontmatter}

\section{Introduction}

ANTARES~\cite{antares} is the world's first deep sea neutrino telescope. The first
detector elements were deployed in March 2006 and data taking started soon after. The
construction was completed by mid-2008.
Until recently, only muons created by muon neutrino charged current ($\nu_\mu$ CC)
interactions in the water that surrounds the detector or in the rock beneath it have been
used in searches for cosmic neutrino sources. Highly energetic muons induce Cherenkov
radiation in the water at a characteristic angle of $\vartheta_\mathrm{Ch} \approx
\unit{42}{\degree}$, which gets recorded by the detector's optical modules. The charge and
timing information of the photon-detections -- referred to as \emph{hits} -- are used to
reconstruct the direction of the parent neutrino with a median angular resolution of
$\xi_\mathrm{track} \approx \unit{0.4}{\degree}$ for an $E^{-2}$ spectrum~\cite{lastPS}.
However, muon tracks constitute only a part of the possible event signatures of
astrophysical neutrinos. Charged current interactions of electron neutrinos ($\nu_e$ CC)
create a shower of electrically charged particles.
All neutrino flavours can interact through neutral current (NC).
In these interactions, only a small fraction of the neutrino energy is transferred to 
a \emph{hadronic shower}. The residual energy is carried away by the neutrino.
Furthermore, tau leptons produced
in $\nu_\tau$ CC interactions decay with a branching ratio of $17\,\%$ into the muon
channel, $65\,\%$ into a hadronic and $18\,\%$ into an electromagnetic
shower.

Due to neutrino oscillation, the cosmic neutrino flux measured at Earth
should constitute a flavour ratio around $\varPhi_{\nu_e} : \varPhi_{\nu_\mu} :
\varPhi_{\nu_\tau} = 1:1:1$~\cite{NuOsciFlux}. 
Especially in the light of the recent discovery of high-energy cosmic neutrinos by the
IceCube experiment, where shower events provided the majority of the neutrino
candidates~\cite{IC2015}, it becomes much more important to increase the sensitivity to
channels that produce particle showers. A major advantage of showers compared to muon
tracks is their inherently low background: The main background for neutrino telescopes is
comprised of tracks by atmospheric muons which are topologically different from showers.
Misidentified muons and electron neutrinos produced by cosmic rays in the upper atmosphere
present the main background in the shower channel. The rate at which electron neutrinos
are produced in the atmosphere at the energy of interest of neutrino telescopes
(\SIrange{1}{1000}{TeV}) is more than a factor of 10 less compared to atmospheric muon
neutrinos.

High-energy muons can travel straight for several kilometres through the rock and water
surrounding the detector.
Showers, on the other hand, deposit all their energy within a few metres from their
interaction vertex. For ANTARES they can be approximated as a point source that emits
light in all directions, though with more intensity at the Cherenkov angle with respect
to the direction of the parent neutrino.
An early reconstruction method for showers has been already used for the search of a
diffuse flux of cosmic neutrinos~\cite{old_showers}. However, the method provided
insufficient angular accuracy for point-source searches compared to the sensitivity level
reached using track events.

In this paper, an algorithm optimised for accurate reconstruction of energy and direction
of shower events in the  ANTARES detector (section~2) is presented. The reconstruction of
the shower position is described in section~3, while the directional and energy
reconstruction and accuracy are presented in section~4. The performances of the method are
discussed in section~5. The results (section~6)  justify adding the selected shower events
to a combined search for neutrino point-sources, as summarised in section~7.

\section{The ANTARES detector}\label{sec:detector}
The ANTARES neutrino telescope is located in the Mediterranean Sea
\unit{40}{\kilo\metre} off the coast of Toulon, France, at \unit{42}{\degree}~%
\unit{48}{\arcminute}~N, \unit{6}{\degree}~\unit{10}{\arcminute}~E. The detector comprises
12 vertical lines anchored at a depth of about \unit{2475}{\metre} and spaced such that
for each line, the closest neighbouring line is located at a distance between about
\unit{60}{\metre}.
Each line is formed by a chain of 25 storeys with an inter-storey distance of
\unit{14.5}{\metre}. Every storey holds 3 optical modules (OMs) housing a single
\unit{10}{\arcsecond} photomultiplier tube (PMT) looking downward at an angle of
\unit{45}{\degree}. The read-out achieves relative time-stamping precision of a nanosecond
between the OMs~\cite{ANTReadOut1, ANTReadOut2}. At the ANTARES site, the transparency and
transmission properties of the sea water~\cite{WaterProp} allow for an excellent timing
measurement of the Cherenkov light induced by relativistic charged particles.

The ANTARES detector has been built in the deep-sea where all daylight is blocked.
However, it is not completely dark in these depths.
Seawater contains the radioactive isotope \textsuperscript{40}K which decays emitting a
relativistic electron. This process produces in each of the ANTARES PMTs a continuous,
ubiquitous background of around \unit{40}{\kilo\hertz}\cite{biolum}.
Additionally, microscopic life forms (mostly bacteria and plankton) are emitting their own
light. This effect is called \emph{bioluminescence} and contributes to the almost constant
baseline rate and also occurs localised in short bursts of a few seconds. These bursts can
cause count rates of several megahertz.

To estimate the reconstruction performance and develop event selection criteria, Monte
Carlo simulations of the different signal and background channels are employed.
Atmospheric muons are simulated using the \emph{MUPAGE} package~\cite{MuPara,mupage},
whereas neutrinos are simulated with the \emph{GENHEN} event generator~\cite{genhen}. The
same sample of simulated events is used for atmospheric and astrophysical neutrinos with
an event-by-event weight to reflect the corresponding neutrino fluxes. For the
atmospheric component, the flux estimate from the Bartol group is used~\cite{bartol}.
The light propagation and the number of photons arriving on the PMTs is
simulated using the \emph{KM3} programme~\cite{simtools1,simtools2} and the
optical background is extracted directly from the data following a run-by-run
approach~\cite{simtools3}.

The longitudinal development of an electromagnetic shower is a well-un\-der\-stood process
governed by the high-energy part of the shower. As described in section 33 of
\cite{pdg_through_matter}, the mean longitudinal profile of the energy deposition in an
electromagnetic shower is reasonably well described by an analytic distribution. This
function is expressed in terms of the scale variable $t=x/X_0$, in which the propagated
distance is measured in units of radiation length $X_0$ ($X_0\approx\unit{36}
{\gram \centi\metre\tothe{-2}}$ for water).
The shape of this distribution was reproduced by our Monte Carlo simulations of electrons
in water. The maximum of the shower lies between about \unit{0.6}{m} (at \unit{1}{GeV})
and \unit{7}{m} (at \unit{100}{PeV}) from the interaction vertex.
Compared to the distances between the OMs in the detector, even the most energetic
showers are compact enough to be approximated by a point-source of light.
Since most charged particles created in the shower propagate roughly towards the original
neutrino direction, most of the photons are still emitted under the Cherenkov angle
$\vartheta_\mathrm{Ch}$ with respect to the parent neutrino direction.
This anisotropy in the number of emitted photons will be exploited to reconstruct the
direction of the shower and thereby to approximate the parent neutrino direction as
described in section~\ref{sec:DirRec}.

\section{Position reconstruction} \label{sec:PosReco}
A proper hit selection is crucial to filter out unwanted background hits caused by the
decay of \textsuperscript{40}K and bioluminescence. For the reconstruction of
the shower position, the subset of hits compatible with a common source of emission, is
identified. Every pair of hits $i, j$ has to fulfil the following causality criterion:
\begin{equation}\label{eq:causality}
    \left|\vec r_i - \vec r_j\right| \geq
    \mathrm c_\mathrm w \cdot \left|t_i - t_j\right|,
\end{equation}
with:\\
\hspace*{\baselineskip}$\vec r_i$, the position of the PMT that recorded hit $i$,\\
\hspace*{\baselineskip}$t_i$, the time at which hit $i$ was recorded and\\
\hspace*{\baselineskip}$\mathrm c_\mathrm w$, the speed of light in water.\\

To understand equation~(\ref{eq:causality}), imagine the position
$\vec r_\mathrm{shower}$ exactly between two PMTs $i$ and $j$.
Their $\Delta r = |\vec r_i - \vec r_j|$ can be arbitrarily high but
$\Delta t = |t_i - t_j|$ is exactly zero. For a generic position
$\vec r_\mathrm{shower}$ and two PMTs close together ($\Delta r$ about $0$), they have to
record their hits at the same time, and thus $\mathrm c_\mathrm w \cdot \Delta t$ must be
small as well. Thus, the time difference between two neighbouring PMTs cannot be
arbitrarily high if they see the same shower, but the time difference between two
arbitrarily distant hits can be zero.
This procedure typically selects between 30 and 60 hits for $\nu_e$ CC interactions from
cosmic neutrinos following an $E^{-2}$ spectrum. Without this hit selection, one would
additionally expect about one hit per OM from the ambient background.
Under the above condition, this common origin of emission -- i.e.\ the shower position
$\vec r_\mathrm{shower}$ and time $t_\mathrm{shower}$ -- can be determined assuming the
following system of quadratic equations:
\begin{equation}\label{eq:pos_rec}
    (\vec r_i - \vec r_\text{shower})^2 =
    \mathrm c^2_\mathrm w \cdot (t_i - t_\text{shower})^2,
\end{equation}
with $1 \leq i \leq N$, where $N$ is the number of selected hits.
The system of equations is linearised by taking the difference between every pair of
equations $i$ and $j$:
\begin{equation}\label{linsys}
    (\vec{r_i}-\vec{r_j}) \cdot \vec{r}_\text{shower} -  (t_i - t_j) \cdot
t_\text{shower}\,\mathrm c_\mathrm w^2
     = \tfrac 1 2 [ |\vec{r}_i|^2-|\vec{r}_j|^2 - \mathrm c_\mathrm w^2 (t_i^2 - t_j^2)]
\end{equation}
for all $i,j:\ 1 \leq i < j \leq N$.
The resulting system of linear equations can be written as:
\begin{equation}\label{eq:matrixform}
    \mathbf A\vec v = \vec b,
\end{equation}
with:\\
\hspace*{\baselineskip}$\vec v = (\vec r_\text{shower}, t_\text{shower})$, the
four-dimensional space-time vector of the shower position,\\
\begin{equation*}
    \mathbf A = \left(\begin{matrix}
        (x_1-x_2) & (y_1-y_2) & (z_1-z_2) & -(t_1-t_2)\mathrm c_\mathrm w\\
        \vdots    &  \vdots   &  \vdots   & \vdots\\
        (x_i-x_j) & (y_i-y_j) & (z_i-z_j) & -(t_i-t_j)\mathrm c_\mathrm w\\
        \vdots    &  \vdots   &  \vdots   & \vdots\\
        (x_{N-1}-x_N) & (y_{N-1}-y_N) & (z_{N-1}-z_N) & -(t_{N-1}-t_N)\mathrm c_\mathrm
w\\
    \end{matrix}\right),
\end{equation*}\\
\begin{equation*}
    \vec b= \frac {1}{2} \cdot \left(\begin{matrix}
        |\vec{r}_1|^2-|\vec{r}_2|^2 - \mathrm c_\mathrm w^2 (t_1^2 - t_2^2)\\
        \vdots\\
        |\vec{r}_i|^2-|\vec{r}_j|^2 - \mathrm c_\mathrm w^2 (t_i^2 - t_j^2)\\
        \vdots\\
        |\vec{r}_{N-1}|^2-|\vec{r}_N|^2 - \mathrm c_\mathrm w^2 (t_{N-1}^2 - t_N^2)
    \end{matrix}\right)
\end{equation*}\\

The matrix $\mathbf A$ has $M=N \cdot (N-1)/2$ rows, therefore
equation~(\ref{eq:matrixform}) represents an over-constrained system of $M$ equations that
can be solved by the method of linear least square fit:
\begin{equation}
    \vec v_\mathrm{l.s.} = (\mathbf A^\text T \mathbf A)^{-1}\mathbf A^\text T \vec b.
\end{equation}

\noindent
A subsequent fit is performed using a robust estimator with the previous fit as starting
point and minimising the so called \emph{M-estimator}, a modified $\chi^2$-like quantity,
defined as:
\begin{equation}
    \label{eq:MEst}
     M_\text{Est} = \sum_{i=1}^{N}
                        \left( q_i \cdot \sqrt{ 1 + {t_{\mathrm{res}\,i}^2} / 2 } \right),
\end{equation}\\*
with $q_i$, the charge of hit $i$ and
\begin{equation} \label{eq:tres}
    t_{\mathrm{res}\,i} = t_i - t_\mathrm{shower} -
            \left|\vec r_i - \vec r_\mathrm{shower} \right| / \mathrm{c}_\mathrm w,
\end{equation}
the \emph{time residual} of hit $i$.\\

\noindent
Like the $\chi^2$ function, $M_\mathrm{Est}$ behaves quadratically for small values of
$t_\mathrm{res}$ but becomes asymptotically linear for larger values. Consequently, it is
less sensitive to outliers, e.g.\ hits from ambient background or scattered photons
which do not fulfil the strict relation in equation~(\ref{eq:pos_rec}). The minimisation is performed by the \textsc{TMinuit2} class
within the ROOT framework~\cite{rootcern}.

\section{Direction and energy reconstruction}\label{sec:DirRec}

The procedure that determines the shower direction (direction fit)  makes use of a second
dedicated hit selection performed on the full set of hits in the event. In particular,
the charges of all hits on a given PMT in a time residual window of
$-200 < t_\mathrm{res}/\si{\nano\second} < 500$ with respect to the already performed
position fit are summed up to yield $q_i$.

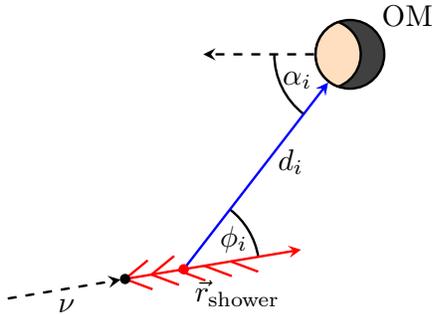
\begin{figure}[h]
    \centering
    \resizebox{.5\textwidth}{!}{
        \begin{tikzpicture}[thick,scale=1.15]
            \tikzset{arcnode/.style={
            decoration={
                        markings, raise = 2mm,
                        mark=at position 0.5 with {
                                    \node[inner sep=0] {#1};
                        }
            },
            postaction={decorate}
            }
            }
            \newcommand*\marktheangle[6]{
            \draw[thick,arcnode={#5}] let \p2=($(#2)-(#1)$),%
                        \p3=($(#3)-(#1)$),%
                        \n2 = {atan2(\y2,\x2)-#6},%
                        \n3 = {atan2(\y3,\x3)}%
                        in ($(\n2:#4)+(#1)$) arc (\n2:\n3:#4);
            }

            \newcommand*\drawwithangle[6][]{ 
                \coordinate (tempbase) at ($(#2)!#4!(#3)$);
                \path let \p2=($(#3)-(#2)$),%
                \n2 = {atan2(\y2,\x2)}%
                in coordinate (coor) at ([shift=(#5+\n2:#6)]tempbase);
                \draw[#1] (tempbase) -- (coor);
            }

            \coordinate (orig) at (0,0);
            \coordinate (nudir) at (3,0.5);
            \coordinate (vert) at ($(0,0)!.4!(nudir)$);
            \coordinate (pos) at ($(0,0)!.6!(nudir)$);
            \coordinate (om) at (3.5,2.5);
            \draw[dashed,-stealth,shorten >= 1.5pt] (0,0) -- (vert);
            \node[anchor=north] at ($(0,0)!.2!(nudir)$) {$\nu$};
            \draw[dashed,-stealth] (om) -- +(-1.5,0) coordinate (dir);
            \draw[fill=darkgray] (om) circle (10pt);
            \node[anchor=south west] at ([shift=(40:8pt)]om) {OM};
            \draw[fill=orange!25!white] ([shift=(250:10pt)]om) arc (110:250:-10pt)
                                                     arc (110:250:+10pt);

            \marktheangle{om}{dir}{pos}{22pt}{$\alpha_i$}{360}
            \marktheangle{pos}{nudir}{om}{22pt}{$\phi_i$}{0}

            \drawwithangle[red]{vert}{nudir}{0}{+30}{.3};
            \drawwithangle[red]{vert}{nudir}{0}{-30}{.3};
            \drawwithangle[red]{vert}{nudir}{0.15}{+30}{.3};
            \drawwithangle[red]{vert}{nudir}{0.15}{-30}{.3};
            \drawwithangle[red]{vert}{nudir}{0.3}{-30}{.3};
            \drawwithangle[red]{vert}{nudir}{0.45}{-30}{.3};
            \drawwithangle[red]{vert}{nudir}{0.6}{-30}{.3};
            \draw[blue,text=black,-stealth,shorten >= 11.9pt]
                        (pos) -- (om) node[midway,right] {$d_i$};
            \draw[-stealth,red] (vert) -- (nudir);
            \fill[black] (vert) circle (1.5pt);
            \fill[red] (pos) circle (1.5pt)
                    node[below right,black]{$\vec r_\mathrm{shower}$};
        \end{tikzpicture}
    }%
    \begin{minipage}[b]{\captwidth}
        \widecapt
        \caption{Geometric variables
            considered by the likelihood function in equation~(\ref{eq:fullLLHood}):
            photon emission angle $\phi_i$, shower--OM distance $d_i$ and photon incident
            angle $\alpha_i$ on the PMT photocathode\textsuperscript{1}.
            \bigskip}
        \label{fig:scheme}
    \end{minipage}%
\end{figure}

A likelihood function is defined to describe the probability $P(q_i)$ that a hypothetical
neutrino $\nu$ with energy $E_\nu$, direction $\vec p_\nu$ and creating a shower at
position $\vec r_\mathrm{shower}$ causes hits with a total measured charge $q_i$ on a PMT
$i$. The measured charge is compared to the expectation value of the number of photons on
this PMT for such a shower. This expectation value depends on the neutrino energy $E_\nu$,
the distance $d_i$ of the OM to the nominal shower position, the photon emission angle
$\phi_i$ from the neutrino direction and its incident angle $\alpha_i$ on the PMT
photocathode\footnote{\label{ft:geom}Note that in case of scattering, a Cherenkov photon
does not travel along the shortest connection between the shower position and the OM,
which defines the distance $d_i$  (cf.\ figure~\ref{fig:scheme}). The angles $\phi_i$ and
$\alpha_i$ are defined w.r.t.\ the shortest connection, irrespective of the actual path of
a scattered photon.}.
A schematic overview of the geometric variables that enter this signal part of the
likelihood function is given in figure~\ref{fig:scheme}. The likelihood also takes into
consideration that the hit could be caused by ambient background and evaluates the
probability that a background event causes a charge as observed on the PMT
($P_\mathrm{bg}(q_i)$). The PMTs that did not record any hits which passed the hit
selection are also taken into account ($P(q_i=0)$).

The likelihood is given by:
\begin{align}\label{eq:fullLLHood}
    \mathscr{L} \hspace{8pt} = \hspace{8pt}
    &\sum_{i=1}^{N} \log\left\{
        P(q_i | E_\nu, d_i, \phi_i, \alpha_i) + P_\text{bg}(q_i) \right\} \nonumber \\
   +&\sum_{i=1}^{N'} \log\left\{
        P(q_i=0 | E_\nu, d_i, \phi_i, \alpha_i) \right\},
\end{align}
with $N$, the number of PMTs with hits, $N'$, the number of PMTs with no hits.

\subsection[The signal term]{The signal term \textendash\
                             $P(q_i | E_\nu, d_i, \phi_i, \alpha_i)$}
The signal term of the likelihood function is determined from a three-dimen\-sional table
obtained from Monte Carlo simulations. It contains, for a given distance between shower
and OM $d_i$, photon-emission angle $\phi_i$ and photon-impact angle $\alpha_i$, the
expectation value of the number of photons on this PMT for a \unit{1}{\tera\electronvolt}
neutrino: $\mathscr N_0(d_i, \phi_i, \alpha_i)$.
The number of emitted photons -- and, therefore, the number $\mathscr N_i$ of
expected photons on the PMT -- is proportional to the neutrino energy. For energies
different from \SI{1}{\tera\electronvolt}, the number of photons is scaled accordingly:
\begin{equation}\label{eq:NScale}
    \mathscr N_i = \mathscr N(E_\nu, d_i, \phi_i, \alpha_i) =
        \mathscr N_0(d_i, \phi_i, \alpha_i) \times E_\nu / \SI{1}{\tera\electronvolt}.
\end{equation}

\noindent
The probability to detect $n$ photons when $\mathscr N$ are expected is given by
the Poisson distribution:
\begin{equation}
    P(n | \mathscr N) = \frac {\mathscr N^n}{n!} \mathrm{e}^{-\mathscr N}.
\end{equation}

\noindent
To first order, the charge $\mathcal Q$  expected to be measured by an ideal PMT is assumed
to be proportional to the number of photons $n$ detected by the PMT:
\begin{equation}
    \mathcal Q = n \times \si{pe},
\end{equation}
with \si{pe} (or \emph{photo-electron}), the average charge measured by the PMT
caused by a single photon.
However, this number of photons $n$ cannot be measured with absolute precision. In
reality, the measured PMT charge $q$ is affected by an uncertainty in form of a
Gaussian centred around the expected charge $\mathcal Q$ with width proportional to
$\sqrt{n}$. For simplicity, this smearing of the charge is approximated by a
continuous extension of the Poisson formula that uses the Gamma function $\Gamma$, defined
for real numbers $q' = q/\si{pe}$:
\begin{equation}
    P(q' | \mathscr N) = \frac{ \mathscr N^{q'} }{ \Gamma( q'-1) } \mathrm{e}^{-\mathscr N}.
\end{equation}

\noindent
Moreover, the read-out electronics saturates at charges above about \unit{20}{pe}
preventing the proper determination of the number of arriving photons for large signals.
For this reason, to obtain a reasonable probability for the measured charge, measured
charges and expected charges above \unit{20}{pe} are treated as being at \unit{20}{pe}.

\subsection[The non-hit term]{The non-hit term \textendash\
                             $P(q_i = 0 | E_\nu, d_i, \phi_i, \alpha_i)$}

The probability to have a non-hit PMT $i$ is simply the Poisson probability to have zero
charge while expecting $\mathscr N_i$ photons to arrive on the photocathode:
\begin{equation}
    P(q_i = 0 | E_\nu, d_i, \phi_i, \alpha_i) = P(q_i=0 | \mathscr N_i) = \mathrm e ^{-\mathscr N_i}.
\end{equation}

\subsection[The background term]{The background term \textendash\ $P_\text{bg}(q_i)$}
\begin{figure}[h]
    \resizebox{\figwidth}{!}{
        \input{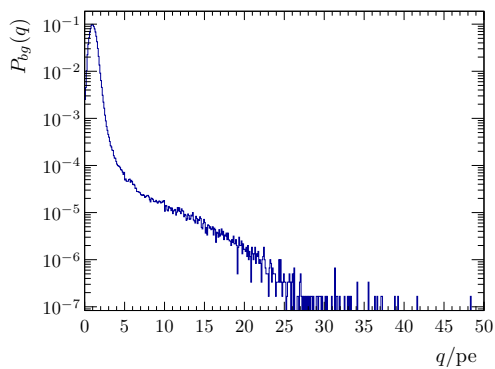}
    }%
    \begin{minipage}[b]{\captwidth}
        \widecapt
        \caption{Unbiased distribution of the charge caused by environmental and
                 atmospheric background. \bigskip}
        \label{fig:BackgroundCharge}
    \end{minipage}%
\end{figure}

The background term gives the probability that one of the uncorrelated background sources
-- explained in section \ref{sec:detector} -- causes the observed charge $q_i$.
Figure~\ref{fig:BackgroundCharge} shows the unbiased distribution of the charge caused by
environmental and atmospheric background. The average value for the shown
distribution is $Q_\mathrm{bg} = \unit{1.1}{pe}$.

\subsection{Implementation}

The energy and direction fit is performed with the \textsc{TMinuit2} class as well. The
probability density function (PDF) used in the fit is provided as a table with discrete
bins.
The minimiser algorithms require the likelihood function to have well defined derivative
at each point. Therefore, the PDF is interpolated with the method of
\emph{trilinear interpolation}.
In order to find the global minimum and avoid possible local minima in the likelihood
landscape, the energy-direction fit has been performed with 12 different starting
directions, corresponding to the directions of the corners of an icosahedron
(as seen from its centre). In the end, the fit with the maximum likelihood value is
selected as the final energy-direction estimate.

\subsection{Error estimator}\label{sec:errest}
The direction fit also provides an angular error estimate $\beta_\mathrm{shower}$ on the
fit direction. After the best direction has been determined, the likelihood landscape
around the fit is scanned along concentric circles of angular distances iteratively
increasing in one-degree steps. The largest angular distance for which the difference
between the likelihood value of any of the test directions and of the best-fit value is
still smaller than $1$ is used as the angular error estimate $\beta_\mathrm{shower}$.

\section{Reconstruction performance}
The performance of the reconstruction algorithm is evaluated applying it to contained
events for which the simulated neutrino interaction vertex lies inside the
instrumented detector volume (horizontal distance from the detector centre
$\rho_\mathrm{MC} < \unit{90}{\metre}$ and vertical distance from the detector centre
$|z_\mathrm{MC}| < \unit{200}{\metre}$). A cut on the angular error estimator was applied
as well ($\beta_\mathrm{shower} < \unit{10}{\degree}$).

\subsection{Position reconstruction}
\begin{figure}[h]
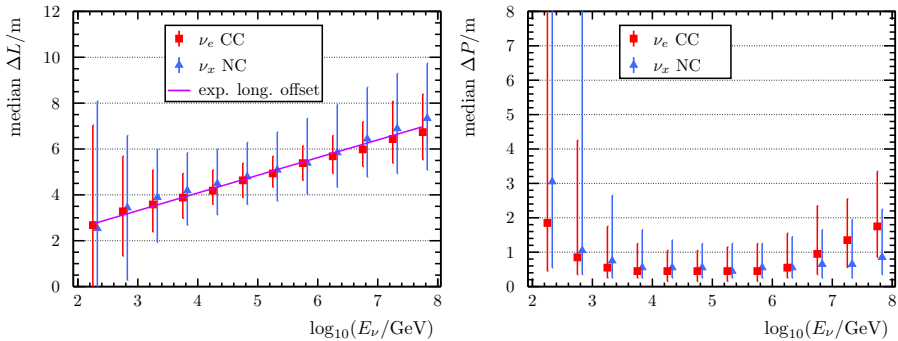

    \resizebox{\textwidth}{!}{
        \input{pics/plots/RecoPerformance/median_DeltaL_emuNC.tex}%
        \input{pics/plots/RecoPerformance/median_DeltaP_emuNC.tex}%
    }%
    \caption{Distance between the true position of the neutrino interaction vertex and the
             reconstructed shower position in the longitudinal (left) and perpendicular
             (right) directions along the neutrino axis. The markers correspond to
             electromagnetic (red) and hadronic (blue) showers after applying the
             containment and angular error cuts ($\rho_\mathrm{MC} < \SI{90}{\metre}$,
             $|z_\mathrm{MC}| < \SI{200}{\metre}, \beta_\mathrm{shower} <
             \unit{10}{\degree}$). The purple line indicates the expected longitudinal
             offset from the neutrino interaction vertex for electromagnetic showers. The
             error bars show the $68\,\%$ spread of the distribution in each energy bin.
    }
    \label{fig:PosPerform}
\end{figure}

Since the reconstruction assumes \emph{one} common point of emission for all photons, it
will most likely reconstruct a position along the shower axis and not the actual neutrino
interaction vertex. Instead, the shower position corresponds to the
intensity weighted mean position of the light emission spectrum for electromagnetic
showers as parametrised in reference~\cite{pdg_through_matter}.
Figure~\ref{fig:PosPerform} shows the longitudinal and perpendicular offset of the shower
position fit with respect to the simulated neutrino interaction vertex. For $\nu_e$ CC and
NC induced showers, the reconstructed position along the shower axis agrees well with the
expected offset from the electromagnetic shower parametrisation.
The median perpendicular distance to the neutrino axis is of the order of half a metre for
both charged and neutral current events over a wide energy range.

\subsection{Direction reconstruction}
\begin{figure}[h]
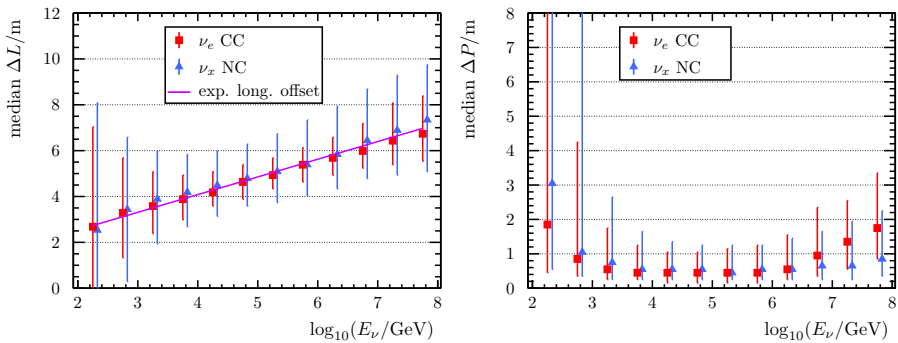

    \resizebox{\textwidth}{!}{
        \input{pics/plots/RecoPerformance/median_DeltaL_emuNC.tex}%
        \input{pics/plots/RecoPerformance/median_DeltaP_emuNC.tex}%
    }%
    \caption{Left: Median angle between the directions of the reconstructed shower and the
            Monte Carlo neutrino as a function of the neutrino energy. Right: Median ratio
            of the reconstructed energy and the Monte Carlo shower energy as a function of
            the Monte Carlo shower energy, i.e.\ the sum of the energy of all mesons and
            charged leptons produced in the initial neutrino interaction. The markers
            show electromagnetic (red) and hadronic (blue) showers after the containment
            and angular error cuts
            ($\rho_\mathrm{MC} < \SI{90}{\metre}$, $|z_\mathrm{MC}| < \SI{200}{\metre}$,
            $\beta_\mathrm{shower} < \unit{10}{\degree}$). The error bars show the
            $68\,\%$ spread of the distribution in each energy bin.
    }
    \label{fig:EnDirPerform}
\end{figure}

The shower angular resolution is defined as the median angle $\xi_\mathrm{shower}$ between
the simulated neutrino and the reconstructed shower directions. As shown in
figure~\ref{fig:EnDirPerform} (left), for contained events and energies in the range
$1 \lesssim E_\nu / \si{\tera\electronvolt} \lesssim 10^3$ it reaches values as low as
\unit{2.3}{\degree} with $16\,\%$ of the events below \unit{1}{\degree}.
For neutrino energies below \unit{1}{\tera\electronvolt}, there is not enough light
produced to illuminate a sufficient number of PMTs for a proper reconstruction. Above
$E_\nu \approx \SI{e3}{\tera\electronvolt}$, the read-out electronics is starting to
saturate and the limited size of the ANTARES detector prevents accessing higher energies
with proper resolutions.

Since only a small fraction of the neutrino energy is transferred to the nucleus in NC
interactions, a hadronic shower created by a high-energy neutrino has correspondingly less
energy than an electromagnetic shower created by an electron neutrino of the same energy
in a CC interaction. For this reason, the angular resolution for hadronic showers above
\unit{e3}{TeV} does not deteriorate as quickly with increasing neutrino energy as for
electromagnetic showers.

\subsection{Energy}
\begin{figure}[h]
    \resizebox{\figwidth}{!}{
        \input{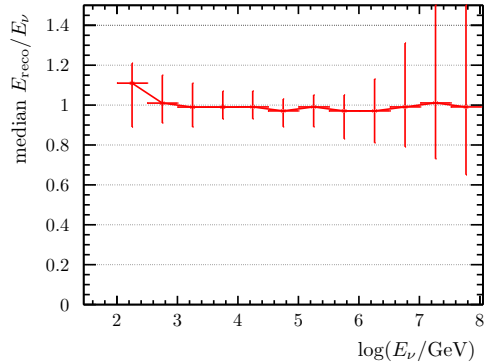}
    }%
    \begin{minipage}[b]{\captwidth}
        \widecapt
        \caption{Ratio between reconstructed energy and MC shower energy for $\nu_e$
                CC events corrected for the bias seen in figure~\ref{fig:EnDirPerform}.
                The performance is shown for $\nu_e$ CC events after the containment and
                angular error cuts ($\rho_\mathrm{MC} < \SI{90}{\metre}$,
                $|z_\mathrm{MC}| < \SI{200}{\metre}$,
                $\beta_\mathrm{shower} < \unit{10}{\degree}$). The error bars show the
                $68\,\%$ spread of the distribution in each energy bin.
                }
        \label{fig:relShE_corrected}
    \end{minipage}
\end{figure}

As shown in figure~\ref{fig:EnDirPerform} (right), a statistical resolution of the shower
energy (which is equal to the neutrino energy only for $\nu_e$ CC events) of
\SIrange{5}{10}{\%} has been achieved. A systematic underestimation of about $20\,\%$ in
the reconstructed energy can be observed over the whole energy range. This effect is
corrected by unfolding the reconstructed energy with the right plot of
figure~\ref{fig:EnDirPerform} so that the median ratio between reconstructed and
\emph{true} Monte Carlo shower energy is flat at $1$ (see
figure~\ref{fig:relShE_corrected}). This energy correction focuses entirely on $\nu_e$ CC
events and does not produce a reliable energy estimate for the neutrino energy in NC
events. The systematic effect of the energy estimation on a combined set comprising
NC and $\nu_e$ CC events can be accounted for in the specific analyses using this method.

\subsection{Angular resolution measured in data}
\begin{figure}[h]
    \resizebox{\figwidth}{!}{
        \input{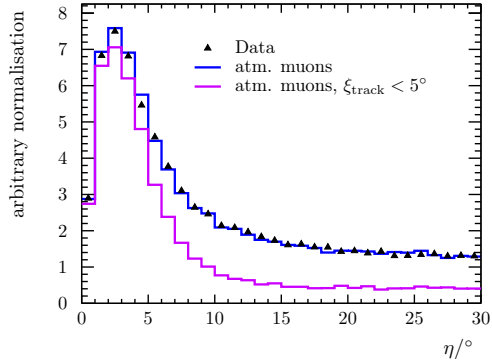}
    }%
    \begin{minipage}[b]{\captwidth}
        \widecapt
        \caption{The angular separation $\eta$ of the reconstructed directions using track
                and shower hypotheses applied to the same atmospheric muon events. The
                angle is shown for data (black), simulated atmospheric muons (blue) and
                simulated atmospheric muons reconstructed with an angular error less than
                \unit{5}{\degree} (purple).\bigskip}
        \label{fig:res_in_data}
    \end{minipage}
\end{figure}

The angular resolution of the shower reconstruction can also be measured directly in data
using a sample of atmospheric muons.
Muons can induce electromagnetic showers through stochastic energy loss processes.
These muon-induced showers will have approximately the same direction as the muon.
As the muon is accurately reconstructed by the track fit, a sample of electromagnetic
showers of known direction can be isolated and the reconstructed shower direction compared
to the direction of the reconstructed muon track.
Figure~\ref{fig:res_in_data} shows the result for a loose selection (i.e.\ containment,
M-estimator, error estimator and GridFit ratio~\cite{EVisser_Thesis} as explained in the
next section and shown in table~\ref{tab:ShowSel}). A clear
population of well reconstructed showers is visible, with a resolution
of two to three degrees (maximum of the distribution). This peak is well modelled in
simulations of atmospheric muons, which implies that the Monte Carlo can
be reliably used to determine the resolution for showers of cosmic origin.
A cut of \unit{5}{\degree} on the angle between the directions of the
simulated and the reconstructed muon has been applied to ensure that the
peak is populated with truly well-reconstructed events.

\section{Event selection and data -- Monte Carlo comparison}

The discrimination of the showers produced by astrophysical neutrinos from the showers
produced by the background of atmospheric muons and neutrinos is a challenging task. The
main limitation is the worse angular resolution compared to muon tracks~\cite{lastPS} and
the fact that muons can also induce electromagnetic showers along their track. The
advantage is the much better energy resolution.

In the following, the performances of the algorithm to discriminate cosmic showers from
the atmospheric background using Monte Carlo observable variables is described. The cosmic
signal is characterised by a power law function of the energy with a harder spectral
index. The results obtained in this section are only illustrative of the methods and can
be adapted according to the specific requirements of different analyses.
In particular, the results of a first all-flavour neutrino point-like source search using
nine years of the ANTARES data are already public~\cite{comb_pssearch}.
The focus in the following is mainly on the reduction of the atmospheric muon
contamination, while maintaining the largest fraction of the cosmic signal.
In particular, the exact cut values have not been optimised (e.g.\ for best sensitivity
or discovery potential in the point-source search).

\begin{figure}[!t]
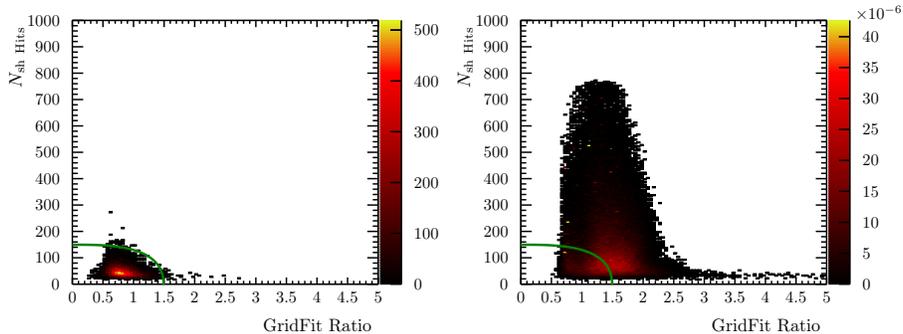

    \resizebox{\linewidth}{!}{%
        \input{pics/plots/MuVeto/muveto_GFR_mupage_precut}%
        \input{pics/plots/MuVeto/muveto_GFR_nueCC_precut}%
    }%
    \caption{Distribution (colour scale on the right for the number of events) of events
            with number of selected hits $N_\mathrm{sh\,hits}$ versus the  GridFit ratio
            $R_\mathrm{GT}$ (see text). The distributions are shown for atmospheric muons
            (left) and for cosmic electron neutrinos undergoing charged current
            interaction creating showers (right) after all previous cuts listed in
            table~\ref{tab:ShowSel}.
            The green line shows the combined $R_\mathrm{GF}$--$N_\mathrm{sh\,hits}$ cut:
            Events below the line are rejected.}
    \label{fig:GFRCut}
\end{figure}

\begin{figure}
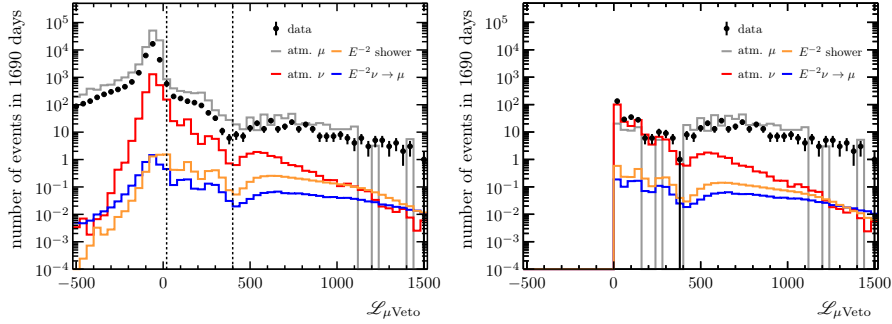

    \centering
    \resizebox{\linewidth}{!}{%
        \input{pics/plots/MuVeto/muveto_RmuVeto_pre}%
        \input{pics/plots/MuVeto/muveto_RmuVeto_post}%
    }%
    \caption{Likelihood muon veto distribution for atmospheric neutrinos (red),
        atmospheric muons (grey), showers caused by astrophysical neutrinos (orange) and
        data (black). The distributions are shown after the GridFit Ratio and all previous
        cuts listed in table~\ref{tab:ShowSel} have been applied (left) and additionally
        after the likelihood-ratio cut (right). The dashed lines mark the cut values:
        Everything below $\mathscr L = 20$ and everything reconstructed as
        $\cos(\vartheta_\mathrm{track}) < -0.2$ and below $\mathscr L = 400$ is rejected.}
    \label{fig:muVetoLike}
\end{figure}

\begin{figure}[h]
    \centering
    \resizebox{.85\textwidth}{!}{
        \input{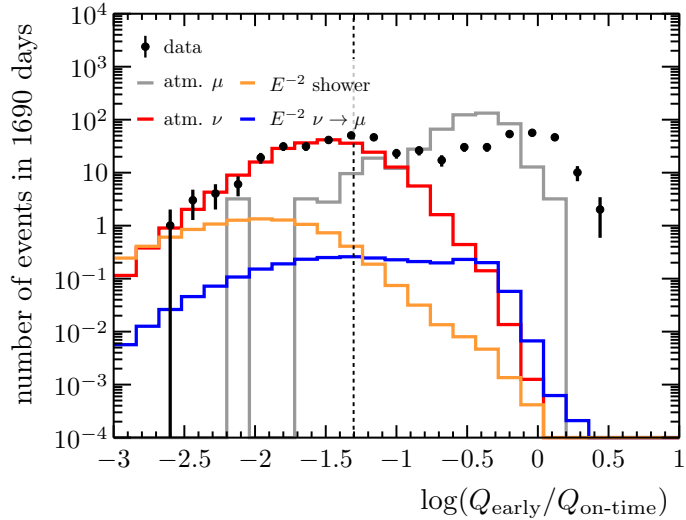}
    }%
    \caption{Distribution of the ratio of the sum of the charges for early and on-time
        hits for atmospheric neutrinos (red), atmospheric muons (grey), showers caused by
        astrophysical neutrinos (orange) and data (black). After the muon veto and all
        previous cuts listed in table~\ref{tab:ShowSel}.
        The dashed line marks the cut value: Everything to the right is rejected.}
    \label{fig:ChargeRatio}
\end{figure}

The effect of the cuts on different Monte Carlo samples are presented in
table~\ref{tab:ShowSel}. The first two columns indicate the name and the value of the
applied condition. Two of them are already presented: the ``up-going'' condition requires
that showers are reconstructed with $\cos(\vartheta_\mathrm{shower}) > -0.1$; the ``error
estimate'' requires that the angular error estimate is $\beta_\mathrm{shower} <
\unit{10}{\degree}$. The other criteria are described in the following. The effect on the
atmospheric muon sample is presented in column 3 ($\epsilon_\mu^\mathrm{atm}$); that on
the atmospheric neutrinos (either yielding a shower or a muon) in
column 4 ($\epsilon_{\nu\rightarrow\mathrm{any}}^\mathrm{atm}$). The effect on a flux of
cosmic neutrinos with spectrum $E^{-2}$ yielding showers of muons simulating a shower are
presented in column 5 ($\epsilon^{E^{-2}}_{\nu\rightarrow\mathrm{shower}}$) and 6
($\epsilon^{E^{-2}}_{\nu\rightarrow\mu}$), respectively.
The last row of the table shows the number of expected events in every channel.
After applying these selection criteria to the ANTARES data
set with an effective life time of 1690 days, 172 shower events remain.
Below, the description of the other criteria applied to reduce the background is presented.
\begin{description}
    \item [Containment+M-Estimator] Reconstructing atmospheric muons with a \linebreak
        shower algorithm often produces ``shower positions'' that lie far away from
        the detector boundary and have a large $M_\mathrm{Est}$ value
        (equation~(\ref{eq:MEst})). A rough selection
        on position and reconstruction quality reduces the amount of background by
        $70\,\%$ already before the direction fit. The quantity $\rho_\mathrm{shower}$ is
        the horizontal distance of the reconstructed shower position from the detector's
        centre and $z_\mathrm{shower}$ is the vertical height above the detector's
        centre.
    \item [Track Veto] To avoid an overlap between the track and shower samples, events
        that pass the muon track selection are excluded from the shower channel.
    \item [GridFit Ratio] The GridFit algorithm was developed for another, recent
        analysis~\cite{EVisser_Thesis}. It is used here to suppress down-going muon
        events. In a first step, it segments the full solid angle in 500 directions. For
        each direction, the number of hits compatible with a muon track from this
        direction is determined. The GridFit ratio $R_\mathrm{GF}$ is the ratio between
        the sum of the compatible hits $N_\mathrm{GFR}$ for all up-going and all
        down-going test directions: $R_\mathrm{GF} = \frac{\sum_\mathrm{up}
        N_\mathrm{GFR}}{\sum_\mathrm{down} N_\mathrm{GFR}}$. A lower value, therefore,
        means a higher likelihood of this event to be a down-going muon. A selection
        criterion combining the GridFit ratio and the number of selected shower hits (see
        figure~\ref{fig:GFRCut}) was devised to further suppress the atmospheric muon
        background.
    \item [Likelihood Muon Veto] In order to improve the discrimination between cosmic
        showers and atmospheric muons, a dedicated likelihood function has been developed.
        This likelihood considers only hits that coincide with another hit on the same
        storey within \unit{20}{\nano\second} and its PDF is based on the following
        parameters:
        \begin{itemize}
            \item time residual $t_\mathrm{res}$ (equation~\ref{eq:tres}) of the hits
                w.r.t. the reconstructed shower position,
            \item number $N$ of on-time hits
                ($-20 < t_\mathrm{res} / \si{\nano\second} < 60$) and 
            \item distance $d$ of the hits to the reconstructed shower position.
        \end{itemize}
        The Likelihood is given by the following equation:
        \begin{equation}\label{eq:muonvetolike}
            \mathscr L_{\mu\mathrm{Veto}} =
                \sum_\mathrm{hits}\Bigl[\log\{P_\mathrm{shower} /
                P_\mathrm{muon}\} + P_\mathrm{shower} - P_\mathrm{muon}\Bigr],
        \end{equation}
        with $P_\mathrm{shower}\ = P(N, d, t_\mathrm{res} | \mathrm{shower})$ and
        $P_\mathrm{muon} = P(N, d, t_\mathrm{res} | \mathrm{muon})$. These PDFs are based
        on the same Monte Carlo simulations mentioned in section~\ref{sec:detector} with
        an energy spectrum proportional to $E^{-2}$ for the cosmic neutrinos that induce
        the showers. The likelihood function shown in equation~(\ref{eq:muonvetolike}) was
        developed to achieve an optimal separation of the shower and muon distributions.
        This likelihood parameter can be combined with the zenith angle, reconstructed by
        the established muon-track fitting algorithm~\cite{lastPS}: On events that have
        been reconstructed as down-going a harder likelihood ratio cut can be applied.
        The distribution for this quantity plotted before and after the combined cut is
        shown in figure~\ref{fig:muVetoLike}. This method further reduces the number of
        atmospheric muons by more than one order of magnitude. Even so, the majority of
        the remaining events consists still of misreconstructed atmospheric muons.
    \item [Charge Ratio] When the shower fit reconstructs a position along the muon track,
        one would expect photons induced by the muon to also arrive earlier than predicted
        by a point source hypothesis. Thus, the charge ratio between the ``early'' and
        ``on-time'' hits was investigated. The distribution of the ratio of those two
        charge-sums is shown in figure~\ref{fig:ChargeRatio}. Here, $Q_\mathrm{early}$ is
        the summed charge of all hits with a time residual of $-1000 \leq t_\mathrm{res} /
        \si{\nano\second} \leq -40$ with respect to the reconstructed shower and
        $Q_\mathrm{on\text{-}time}$ is the summed charge of all hits with time residuals
        of $-30 \leq t_\mathrm{res} / \si{\nano\second} \leq 1000$.
\end{description}%

After reducing the amount of atmospheric muons by six orders of magnitude, just before
the charge-ratio cut (see figure~\ref{fig:ChargeRatio}), the Monte Carlo simulations of
atmospheric muons do no longer well describe the data in the right part of the plot.
The discrepancy lies well out of the acceptance region wherein the data agrees with
the simulation of atmospheric neutrino events.

The event selection does not only reject unwanted background events but also poorly
reconstructed signal events. The direction resolution improves slightly compared to what
is shown in figure~\ref{fig:EnDirPerform}, particularly in the lower energy region.

\begin{table}[!h]
    \centering
    \caption{Event selection criteria for the shower channel and the selection efficiency
        after each step for atmospheric muons and neutrinos and cosmic neutrinos creating
        a shower in the detector. The efficiencies are defined as the ratio of the number
        of events that passed a cut and the number of events after the trigger selection.
        In the last row, the number of events expected from the simulation is shown.
        \medskip}
    \label{tab:ShowSel}
    \centering
    \resizebox{\linewidth}{!}{
        \begin{tabular}{llrrrr}
            Criterion      & Condition
                & $\epsilon^\mathrm{atm.}_\mu$
                & $\epsilon^\mathrm{atm.}_{\nu\rightarrow\mathrm{any}}$
                & $\epsilon^\mathrm{E^{-2}}_{\nu\rightarrow\mathrm{shower}}$
                & $\epsilon^\mathrm{E^{-2}}_{\nu\rightarrow\mu}$  \medskip\\

            Triggered      &
                & $100 \,\%$ & $100 \,\%$ & $100 \,\%$ & $100 \,\%$ \\
            Containment    & $\rho_\mathrm{shower} < \unit{300}{\metre}$,
                             $|z_\mathrm{shower}| < \unit{250}{\metre}$
                & $ 53 \,\%$ &  $81 \,\%$ & $ 93 \,\%$ &  $75 \,\%$ \\
            M-Estimator    & $M_\mathrm{Est} < 1000$
                & $ 40 \,\%$ & $ 66 \,\%$ & $ 90 \,\%$ &  $72 \,\%$ \\
            Track Veto     & not selected as muon candidate
                & $ 40 \,\%$ & $ 59 \,\%$ & $ 88 \,\%$ &  $49 \,\%$ \\
            Up-Going       & $\cos(\vartheta_\mathrm{shower}) > -0.1$
                & $ 18 \,\%$ & $ 44 \,\%$ & $ 46 \,\%$ &  $28 \,\%$ \\
            Error Estimate & $\beta_\mathrm{shower} < \unit{10}{\degree}$
                & $0.66 \,\%$ &$  5.0 \,\%$&$ 26  \,\%$&$ 9.3 \,\%$ \\\smallskip
            GridFit Ratio  & $\left(\frac {R_\mathrm{GF}}{1.3}\right)^3
                            + \left(\frac{N_\mathrm{sh\ hits}}{150}\right)^3 > 1$
                &$0.057 \,\%$ &$ 4.2 \,\% $&$ 22  \,\%$&$ 6.1 \,\%$ \\
            Muon Veto      & $\mathscr L_\mathrm{\mu Veto} >
                              \begin{cases}
                                  400, & \text{if } \cos(\vartheta_\mathrm{track})
                                                    < -0.2\\
                                  20 , & \text{otherwise}
                              \end{cases}$
                & \num{2.9e-4} \,\% & $0.41 \,\%$&$ 17  \,\%$&$ 2.8 \,\%$ \\

            Charge Ratio   & $\log(Q_\mathrm{early} / Q_\mathrm{on\text{-}time}) < -1.3$
                & \num{1.1e-5} \,\% & $0.31 \,\%$&$ 16  \,\%$&$ 1.3 \,\%$\\[0.25em]
            \midrule
            \multicolumn{2}{l}{ expected Events in 1690 days}
                & 18.8 & 163 & 2.78 & 0.63
        \end{tabular}
    }
\end{table}

\section{Conclusion}
An algorithm to reconstruct underwater particle showers has been developed.
It makes use of the fact that the highly energetic, electrically charged particles induce
Cherenkov radiation mainly under one specific angle with respect to the direction of the
parent neutrino and that this emission profile is preserved over large distances due to
the good optical properties of the deep-sea water. The showers can be approximated as
point sources of photons which emit their light at one given time. The shower position is
reconstructed with a precision of about \SI{1}{\metre} and for the neutrino direction
resolutions of \SIrange{2}{3}{\degree} are achieved.
A statistical uncertainty for the shower energy of about \SIrange{5}{10}{\%} is obtained.

Despite their much more compact event signature, the shower algorithm's angular resolution
is only about a factor of five worse than that of the muon channel. Combined with their
inherently low background, shower events will prove very valuable in the search for
point-like and extended neutrino sources.
Our studies~\cite{comb_pssearch} showed that the shower channel contributes about 23\,\%
of all signal events for an $E^{-2}$ energy spectrum corresponding to an increase of the
point-source sensitivity of about \unit{10}{\%}. The sensitivity of the shower channel to
the ANTARES searches for a diffuse flux of cosmic neutrinos is almost equivalent to that
of the muon channel~\cite{old_showers, antares_diffuse}, due to the better energy estimate
and the lower atmospheric background.

\section{Acknowledgements}
The authors acknowledge the financial support of the funding agencies:
Centre National de la Recherche Scientifique (CNRS), Commissariat \`a
l'\'ener\-gie atomique et aux \'energies alternatives (CEA),
Commission Europ\'eenne (FEDER fund and Marie Curie Program),
Institut Universitaire de France (IUF), IdEx program and UnivEarthS
Labex program at Sorbonne Paris Cit\'e (ANR-10-LABX-0023 and
ANR-11-IDEX-0005-02), Labex OCEVU (ANR-11-LABX-0060) and the
A*MIDEX project (ANR-11-IDEX-0001-02),
R\'egion \^Ile-de-France (DIM-ACAV), R\'egion
Alsace (contrat CPER), R\'egion Provence-Alpes-C\^ote d'Azur,
D\'e\-par\-tement du Var and Ville de La
Seyne-sur-Mer, France;
Bundesministerium f\"ur Bildung und Forschung
(BMBF), Germany;
Istituto Nazionale di Fisica Nucleare (INFN), Italy;
Stichting voor Fundamenteel Onderzoek der Materie (FOM), Nederlandse
organisatie voor Wetenschappelijk Onderzoek (NWO), the Netherlands;
Council of the President of the Russian Federation for young
scientists and leading scientific schools supporting grants, Russia;
National Authority for Scientific Research (ANCS), Romania;
Mi\-nis\-te\-rio de Econom\'{\i}a y Competitividad (MINECO):
Plan Estatal de Investigaci\'{o}n (refs. FPA2015-65150-C3-1-P, -2-P and -3-P, (MINECO/FEDER)),
Severo Ochoa Centre of Excellence and MultiDark Consolider (MINECO),
and Prometeo and Grisol\'{i}a programs (Generalitat Valenciana), Spain;
Ministry of Higher Education, Scientific Research and Professional Training, Morocco.
We also acknowledge the technical support of Ifremer, AIM and Foselev Marine
for the sea operation and the CC-IN2P3 for the computing facilities.


\section{Bibliography}
\bibliographystyle{unsrt}
\bibliography{Paper_ShowerReco}

\begin{thebibliography}{10}

\bibitem{antares}
{The ANTARES Collaboration}.
\newblock {ANTARES: the first undersea neutrino telescope}.
\newblock {\em Nuclear Instruments and Methods}, A656:11, 2011.

\bibitem{lastPS}
{The ANTARES Collaboration}.
\newblock {Searches for Point-like and extended neutrino sources}.
\newblock {\em The Astrophysical Journal Letters}, 786:L5, 2014.

\bibitem{NuOsciFlux}
{J. G. Learned and S. Pakvasa}.
\newblock
  {\href{http://www.sciencedirect.com/science/article/pii/0927650594000433}{Detecting
  $\nu_\tau$ oscillations at PeV energies}}.
\newblock {\em {Astroparticle Physics}}, 3, 1995.

\bibitem{IC2015}
{The IceCube Collaboration}.
\newblock {Atmospheric and astrophysical neutrinos above 1 TeV interacting in
  IceCube}.
\newblock {\em Physical Review D}, 91:022001, 2015.

\bibitem{old_showers}
{The ANTARES Collaboration}.
\newblock An algorithm for the reconstruction of high-energy neutrino-induced
  particle showers and its application to the {ANTARES} neutrino telescope.
\newblock {\em The European Physical Journal C}, 77(6):419, 2017.

\bibitem{ANTReadOut1}
{The ANTARES Collaboration}.
\newblock {Performance of the front-end electronics of the ANTARES neutrino
  telescope}.
\newblock {\em Nuclear Instruments and Methods A}, 622:59, 2010.

\bibitem{ANTReadOut2}
{The ANTARES Collaboration}.
\newblock {Time calibration with atmospheric muon tracks in ANTARES}.
\newblock {\em Astroparticle Physics}, 78:43, 2016.

\bibitem{WaterProp}
{The ANTARES Collaboration}.
\newblock {Transmission of light in deep sea water at the site of the Antares
  neutrino telescope}.
\newblock {\em Astroparticle Physics}, 23:131, 2005.

\bibitem{biolum}
{C. Tambirini, M. Canals, X. Durrieu de Madron, L. Houpert, D. Lefèvre et al.}
\newblock {Deep-Sea Bioluminescence Blooms after Dense Water Formation at the
  Ocean Surface}.
\newblock {\em {PLoS ONE}}, {8(7): e67523}, 2013.

\bibitem{MuPara}
Y.~Becherini et~al.
\newblock A parametrisation of single and multiple muons in the deep water or
  ice.
\newblock {\em Astroparticle Physics}, 25, 2006.

\bibitem{mupage}
G.~Carminati, A.~Margiotta, and M.~Spurio.
\newblock {Atmospheric MUons from PArametric formulas: A Fast GEnerator for
  neutrino telescopes (MUPAGE)}.
\newblock {\em {Computer Physics Communications}}, 179, 2008.

\bibitem{genhen}
D.~Bailey.
\newblock {\em {Monte Carlo tools and analysis methods for understanding the
  ANTARES experiment and predicting its sensitivity to Dark Matter}}.
\newblock PhD thesis, Wolfson College, Oxford, 2002.

\bibitem{bartol}
{V. Agrawal, T. K. Gaisser, P. Lipari and T. Stanev}.
\newblock
  \href{http://journals.aps.org/prd/abstract/10.1103/PhysRevD.53.1314}{Atmospheric
  neutrino flux above \unit{1}{GeV}}.
\newblock {\em Physical Review D}, 53, 1996.

\bibitem{simtools1}
J.~Brunner.
\newblock {ANTARES} simulation tools.
\newblock {\em Proceedings of the VLVnT}, 2003.
\newblock http://www.vlvnt.nl/proceedings.pdf.

\bibitem{simtools2}
A.~Margiotta.
\newblock Common simulation tools for large volume neutrino detectors.
\newblock {\em Nuclear Instruments and Methods}, 725, 2013.

\bibitem{simtools3}
L.~A. Fusco and A.~Margiotta.
\newblock The run-by-run monte carlo simulation for the {ANTARES} experiment.
\newblock {\em EPJ Web Conference}, 116, 2016.

\bibitem{pdg_through_matter}
{C. Patrignani et al. [Particle Data Group]}.
\newblock {\href{http://pdg.lbl.gov/}{2015 Review of Particle Physics --
  Passage of particles through matter}}.
\newblock {\em {Chinese Physics C}}, 40, 100001, 2016.

\bibitem{rootcern}
{The ROOT Data Analysis Framework}.
\newblock \url{www.root.cern.ch}.

\bibitem{EVisser_Thesis}
E.~L. Visser.
\newblock {\em
  \href{http://www.nikhef.nl/pub/services/newbiblio/theses.php}{Neutrinos from
  the Milky Way}}.
\newblock PhD thesis, Nikhef, 2015.

\bibitem{comb_pssearch}
{The ANTARES Collaboration}.
\newblock {First all-flavour Neutrino Point-like Source Search with the ANTARES
  Neutrino Telescope}.
\newblock arXiv:1706.01857.

\bibitem{antares_diffuse}
{The ANTARES Collaboration}.
\newblock {Search for a diffuse flux of cosmic neutrinos with the ANTARES
  telescope}.
\newblock {\em Proceedings of Science -- ICRC}, 2017.

\end{thebibliography}

\end{document}